# Open Educational Resources: Barriers and Open Issues

Pedro Henrique Dias Valle[a], Rafael Capilla[b], Vinicius dos Santos[a], Daniel Feitosa[b], Elisa Yumi Nakagawa[a]

[a]University of São Paulo, Brazil; [b]Rey Juan Carlos University, Spain;

[c]University of Groningen, The Netherlands

**ABSTRACT**

Open Educational Resources (OER) are freely available teaching and learning materials, such as textbooks, videos, and interactive games, that can be used, reused, adapted, and shared. OER can leverage access, collaboration, and innovation in education; however, their adoption and long-term use remain limited. Motivated by this issue, this manuscript examined the literature and identified 26 social, economic, and technical barriers that hinder teachers, students, and institutions from creating, using, and maintaining OER. These barriers were evaluated through semi-structured interviews with experts to ensure their understandability, correctness, completeness, and relevance. We adopted a four-step research method: (1) a tertiary study that identified barriers from 26 secondary studies; (2) analysis and classification of the barriers according to social, economic, and technical dimensions and the OER lifecycle activities they affect; (3) design of the conceptual model to represent relationships among OER elements, barriers, and mitigation actions; and (4) evaluation through expert interviews. These barriers are also illustrated using a real-world OER. The findings provide insights for the education community, supporting inclusive strategies, institutional actions, and public policies to reduce social, economic, and technical barriers to OER. By addressing factors that affect accessibility, sustainability, and equitable participation, this manuscript advances universal access to educational resources and fosters more inclusive educational ecosystems.

## 1. Introduction

The scientific community and teachers have observed significant technological advances in recent years. At the same time, researchers have considered that those born after this emergence are better qualified to handle this progress [1]. In particular, Prensky [2] referred to such people as *digital natives* due to their ease of interaction with new technologies. At the same time, Adedoyin and Soykan [3] noticed how students' learning has evolved over the years. Different movements have emerged to address changes in learning. One important initiative is *open education*, which combines effective pedagogical practices among teachers with a digital culture of collaboration and interactivity [4]. Such initiative is based on the freedom to use, change, mix, and

CONTACT Pedro Henrique Dias Valle. Email: pedrohenriquevalle@usp.br
These authors contributed equally to this work.

redistribute educational resources through open technologies, prioritizing free software and open formats [5].

It is worth noting that Open Educational Resources (OER) are one of the most important artifacts of open education [4, 6]. OER refers to teaching, learning, and research materials available in any media (digital or otherwise) with a public domain or that have been released using an open license that permits no-cost access, use, adaptation, and redistribution by others with no or limited restrictions [5]. OER has provided several benefits, including [7–9]: (i) providing personalized content; (ii) supporting students with disabilities to learn more effectively and become motivated to learn new content; (iii) providing extra content to overcome difficulties in learning content in the classroom; (iv) providing cost-effectiveness; and (v) allowing teachers to work openly and flexibly to address students' individualized needs. These resources can also enhance collaboration among coworkers and facilitate critical reflections on teachers' pedagogical practices [5]. Furthermore, OER can contribute to a more inclusive and accessible education that can adapt to challenges and societal changes.

OER has been concentrated in developed countries, mainly due to technological infrastructure, institutional incentives, and established policies supporting knowledge sharing. Pioneering initiatives such as MIT OpenCourseWare, launched in 2002, triggered the global open education movement and inspired universities to make their educational materials freely available [10, 11]. Moreover, organizations such as UNESCO and the OECD highlight that the success of OER in countries like the United States, the United Kingdom, and Canada is strongly related to the presence of public policies, funding programs, and a collaborative culture among educators and institutions [12]. In contrast, in developing countries, the absence of government strategies and low investment in digital infrastructure continue to limit the dissemination and effective use of these resources [13].

Despite the benefits and some propagation of OER, Otto et al. [14] observed that most OER have not been systematically used for a long time. Some researchers associate the low (re)use with the lack of maintenance over the years [15, 16]. In turn, this lack may be mainly due to the cost of producing content, the low incentives to share, or the lack of evidence of the OER effectiveness [15–18]. The scientific community has consistently debated the need to develop products or artifacts that remain in use over the years [19, 20], while few studies have investigated the longevity of OER. In particular, Carvalho et al. [15] discussed the costs of maintaining and updating existing OER. These costs relate to delivery channels, adapting the OER to changes in the support platforms, localizing OER, and producing new content. Complementarily, Tlili et al. [21] investigated the potential of OER models, often implemented by universities and consortia, to sustain these resources' continued use. In this scenario, investigating how to reduce costs, promote pedagogical innovation, foster global collaboration, and minimize environmental impact during OER development and (re)use is essential.

The main objective of this work is to present the key barriers that have hindered the effective creation, adoption, and long-term utilization of OER. To achieve this, we conducted an extensive literature review to identify and analyze the barriers. Following this, we categorized the 26 resulting barriers according to: (i) their nature (i.e., social, economic, and technical aspects); and (ii) the activities of the OER life cycle that the barriers impact (e.g., creation, use, reuse, and redistribution). We then evaluated these barriers through semi-structured interviews with experts to ensure their understandability, correctness, completeness, and relevance. Additionally, we illustrated how these barriers manifest in a real-world OER example. As a key outcome, this work highlights the need for the education community to better understand and address the challenges that prevent teachers, students, and institutions from fully benefiting from OER, ultimately supporting broader access and equity in education.

The remainder of this work is organized as follows: Section 2 presents the background on OER. Section 3 describes the research method. Section 4 details our evaluation. Section 5 presents the barriers that hinder the use of OER. Section 6 presents and discusses the implications of our findings for the education community, highlighting their potential impact on educational practice, institutional strategies, and policy-related decisions. Section 7 examines the main future actions and the threats to the validity of this work, discussing potential limitations, mitigation strategies, and directions for future research. Finally, Section 8 outlines the final remarks.

## 2. Background

This section outlines the key concepts related to OER and their characteristics. OER originated from the current discourse on the digitalization of education [14], but the term *Open Educational Resource* (or simply OER) has existed for more than 20 years [22, 23]. UNESCO coined this term in 2002 during the forum on the *Impact of Open Courseware for Higher Education in Developing Countries* [24]. According to UNESCO [16], OER comprises learning, teaching, and research materials in any format and media that are in the public domain or are copyrighted under an open license that permits no-cost access, reuse, repurposing, adaptation, and redistribution.

In 2002, MIT established the *OpenCourseWare Project*, which provided free access to several classes and enabled users to modify and share them for non-commercial purposes [25]. Next, several initiatives emerged worldwide to disseminate OER to the education community [26]. However, such actions come with a cost. Among the challenges identified for expanding the creation, use, and distribution of OER, the following stand out [25]: (i) Most teachers are unaware of the existence of OER; (ii) Governments and educational institutions are also unaware of the existence of OER or are not convinced of the benefits provided by open education; and (iii) Different types of licenses confuse or even create incompatibility in using OER.

Hence, public policies should be implemented to encourage and educate teachers on proper creation processes and on fostering collaborative work within a participatory culture [11]. For that, research is also needed to investigate and evaluate existing and new mechanisms that can (further) support OER.

A first and paramount step towards more standardized and inclusive OER is to use more flexible, *copyleft* licenses, such as *Creative Commons*. When a resource is licensed under a copyleft license, the following five activities, also called 5R's, are ideally permitted [26]:

- **Reuse:** Content may be used in its original format without being altered. In addition, it is allowed to use this content on different occasions, such as classes, study groups, websites, and videos;
- **Retain:** Permission is granted to retain copies of the content for personal purposes, allowing the right to own and control copies of the resource, such as downloading, duplicating, storing, and managing the content;
- **Revise:** Content may be adapted, modified, adjusted, and altered to meet specific users' objectives. An example is translating the language of the resource content to enable understanding it in a new language;
- **Remix:** Content may be combined with similar content to create something

new. An example is the incorporation of the content into a mashup; and
- **Redistribute:** Content can be shared with anyone in its original or altered format. Hence, the right to share copies of the original content, revisions, or remixes of resources is allowed.

It is worth noting that benefiting effectively from OER is not a trivial task. There are also several barriers to achieving that. A barrier is any factor that hinders the continuous creation, adoption, adaptation, or maintenance over time. Such barriers may emerge from different contexts (human, technical, or organizational) and can directly affect the longevity, quality, and accessibility of OER.

## 3. Research Method

Our study encompassed a four-step research method, as shown in Figure 1 and described below. We first conducted a tertiary study based on guidelines proposed by Kitchenham et al. [27].

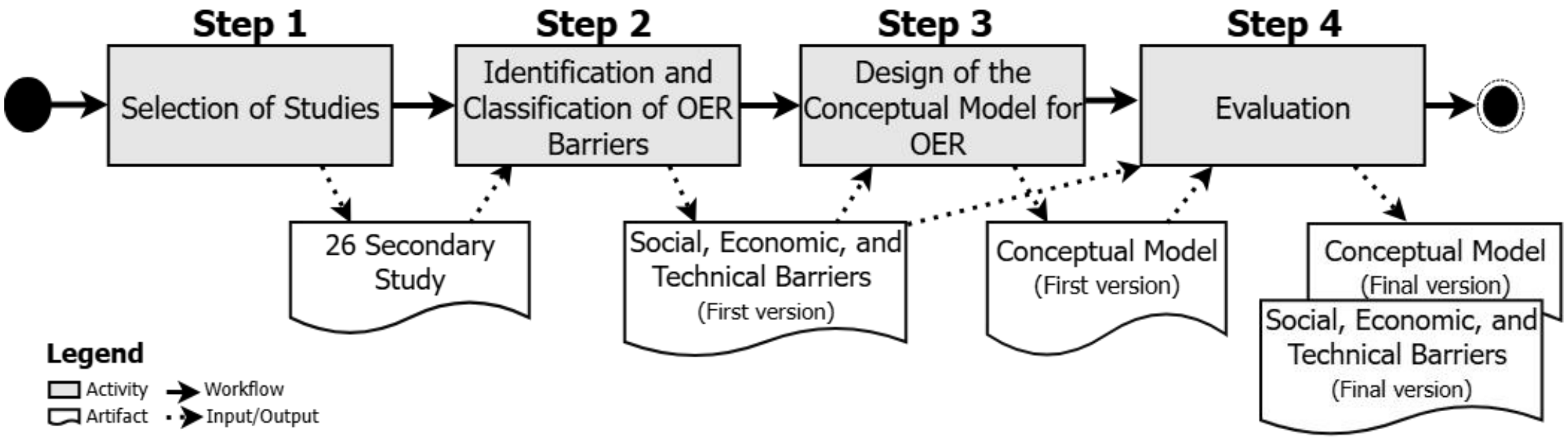

**Figure 1.** Research method

**Step 1: Selection of Studies**

Specifically, our objective was to gather evidence from the scientific literature on the main barriers that teachers and students have encountered in developing, using, reusing, selecting, retaining, remixing, and redistributing OER. We searched for secondary studies published up to 2026. The search was conducted on 14 January 2026 across three digital libraries: Scopus[1] , IEEE Xplore[2] , and the ACM Digital Library[3] , using the following search string *((“literature review” OR “systematic mapping” OR “systematic review” OR “mapping study” OR “systematic map”) AND (“open educational resource” OR “OER”))*. To support the study’s selection, we considered one inclusion criterion (IC) and four exclusion criteria (EC):

- IC1: Secondary studies that address barriers for developing, reusing, selecting, retaining, remixing, and redistributing OER;
- EC2: Studies that are not secondary studies;
- EC1: Secondary studies that do not mention OER barriers;
- EC3: Secondary studies written in languages other than English; and
- EC4: Secondary studies published as short papers.

[1]https://www.scopus.com
[2]https://ieeexplore.ieee.org
[3]https://dl.acm.org

We first configured the search string for each database engine. During the search process, time limits were not imposed, and filters were not applied to title, abstract, or keywords, with the intention of being as inclusive as possible. Upon completing this step, we obtained 389 studies; after removing duplicate studies, we were left with 350. After that, we applied the selection criteria by reviewing the titles, abstracts, and keywords of each study and selected 53 studies. After reading the full texts of each study, we re-applied the criteria and identified 26 secondary studies, listed in Table 1.

**Step 2: Identification and Classification of OER Barriers**

We scrutinized the 26 studies to extract the barriers to OER. We also carefully read the primary studies cited by the selected secondary studies to find barriers. Following this, we removed duplicate barriers, leaving 26 unique barriers. We also classified them according to their nature into three main aspects:

- **Social:** It involves human and organizational factors, such as resistance to adopting open practices, lack of awareness about OER benefits, insufficient collaboration among educators, and limited institutional or cultural support for sharing materials;
- **Technical:** It concerns issues with tools, platforms, and standards, including format incompatibility, poor interoperability, a lack of metadata or accessibility, and difficulties in maintaining or updating OER over time; and
- **Economic:** It relates to the financial and resource constraints that affect the creation and maintenance of OER, such as limited funding, the absence of incentives, and a lack of sustainable business or institutional models.

After this classification, we analyzed each barrier to identify which 5R's activities it could affect. The initial version of the barriers, their classification, and the associated 5R's activities are presented in the external material[4]. Section 5 then presents the final version of them, refined after the evaluation process.

**Step 3: Design of the Conceptual Model on Barriers for OER**

In this step, we developed a conceptual model, named CM4OER, to better explain the barriers to OER. To design it, we identified the main concepts related to OER and barriers and analyzed their relationships. Following this, we created a visual representation of our model. The model incorporates key elements associated with OER (e.g., licenses and types of OER, such as videos and slides), as well as their operations (e.g., redistribution and retention). More importantly, it encompasses the concepts of OER barriers and mitigation actions to address such problems, ensuring the effective use of OER. We compiled these actions from the studies and discussed them in Section 7, along with the open issues for future work. The initial version of the conceptual model is also presented in the external material. Section 5 presents the final version of this model, which has been refined following the evaluation process.

**Step 4: Evaluation**

In this step, we evaluated the barriers and the conceptual model through semi-structured interviews conducted with eight experts. To select participants, we used convenience sampling from the authors' academic and professional networks. Additionally, we demonstrate how a selected real-world OER has addressed these barriers. Section 4 details this evaluation. Following this, we refined the barriers and the conceptual model based on the collected data and insights from the evaluation. Section 5

[4]Available at: https://encurtador.com.br/eqcxK

**Table 1.** List of Secondary Studies

| ID | Title | Year | Reference |
|---|---|---|---|
| S1 | Open educational practices in higher education: Institutional adoption and challenges | 2013 | [28] |
| S2 | Exploration of open educational resources in non-English speaking communities. | 2013 | [29] |
| S3 | Investigating perceived barriers to the use of open educational resources in higher education in tanzania. | 2014 | [30] |
| S4 | A barrier framework for open E-Learning in public administrations | 2015 | [31] |
| S5 | Open educational resources repositories literature review – Towards a comprehensive quality approaches framework | 2015 | [32] |
| S6 | Incentives and barriers to OER adoption: A qualitative analysis of faculty perceptions | 2016 | [33] |
| S7 | Institutional and technological barriers to the use of open educational resources (OERs) in physiology and medical education | 2017 | [34] |
| S8 | MOOCs and OER in the global south: Problems and potential | 2018 | [13] |
| S9 | Cost, outcomes, use, and perceptions of open educational resources in psychology: A narrative review of the literature | 2019 | [35] |
| S10 | Doing MOOCs in Dili: Studying online learner behaviour in the Global South | 2019 | [36] |
| S11 | Designing open informational ecosystems on the concept of open educational resources | 2020 | [37] |
| S12 | The evolution of sustainability models for open educational resources: Insights from the literature and experts | 2020 | [21] |
| S13 | The power of open: benefits, barriers, and strategies for integration of open educational resources | 2020 | [38] |
| S14 | Looking back before we move forward: A systematic review of research on open educational resources | 2020 | [39] |
| S15 | Developing institutional open educational resource repositories in Vietnam: Opportunities and challenges | 2021 | [40] |
| S16 | Quality models and quality attributes for open educational resources: A systematic mapping | 2021 | [41] |
| S17 | Trends and gaps in empirical research on open educational resources (OER): A systematic mapping of the literature from 2015 to 2019 | 2021 | [14] |
| S18 | Open education resources' benefits and challenges in the academic world: a systematic review | 2022 | [42] |
| S19 | Open Educational Resources (OERs) at European Higher Education Institutions in the Field of Library and Information Science during COVID-19 Pandemic | 2023 | [43] |
| S20 | The Affordability Solution: a Systematic Review of Open Educational Resources | 2023 | [44] |
| S21 | Equity, diversity, and inclusion in open education: A systematic literature review | 2023 | [45] |
| S22 | A Systematic Review of Systematic Reviews on Open Educational Resources: An Analysis of the Legal and Technical Openness | 2023 | [46] |
| S23 | Guidelines: The Do's, Don'tsand Don't Knows of Creating Open Educational Resources | 2023 | [47] |
| S24 | Pedagogical and accessibility guidelines for open educational resources focusing on blind students | 2024 | [48] |
| S25 | Factors affecting the sustainability of open educational resource initiatives in higher education: A systematic review | 2025 | [49] |
| S26 | A Review of Reviews on Open Educational Resources | 2025 | [50] |

then presents the final version of the barriers and the conceptual model.

## 4. Evaluation

This section presents two studies carried out to evaluate the identified barriers and CM4OER. To achieve this, we conducted semi-structured interviews and a demonstration of the barriers in practice using an OER identified and selected from a well-known repository.

### *4.1. Interviews*

Interviews are a qualitative research method to capture participants' experiences, perceptions, and interpretations of a phenomenon [51]. Semi-structured interviews, in particular, are beneficial for balancing comparability across respondents with the flexibility to explore emerging insights [52]. Interviews can also enhance the credibility, applicability, and external validity of artifacts, such as frameworks, models, or barrier sets [53]. In evaluation studies (as in our case), interviews help triangulate evidence from the literature by confronting theoretical constructs with practitioners' or experts' real-world perspectives. Hence, we considered that interviews are a suitable method for our evaluation.

This manuscript involved voluntary semi-structured interviews with experts to evaluate the identified barriers and the CM4OER model. Participants were informed about the research goals and provided verbal consent before recording and transcription. No sensitive or personally identifiable data were collected, and all transcripts were anonymized and used only for aggregated qualitative analysis. The study did not involve vulnerable populations, interventions, or risks to participants. According to the Brazilian National Health Council Resolution No. 510/2016, research in the Human and Social Sciences based on expert consultation and opinion that ensures anonymity and does not collect identifiable personal data may be exempt from formal submission to a Research Ethics Committee; therefore, ethics approval was not required.

The rest of this section describes the research objectives, criteria for participant selection, data collection procedures, data analysis methods, and the main findings derived from the interviews.

#### *4.1.1. Interview Design*

The **objective of the interviews** was to evaluate the barriers previously identified in the literature. More specifically, we analyzed: (i) whether participants recognized these barriers as relevant in light of their own experiences; (ii) confirm the adequacy of the classification of the barriers within the three aspects (i.e., social, economic, and technical); and (iii) assess the alignment of the barriers with the five 5R's OER activities: retain, reuse, revise, remix, and redistribute. The objective was also to evaluate CM4OER, which represents the relationship between the barriers and OER. This objective aimed to strengthen the empirical foundation and practical relevance of both the barriers and the conceptual model.

Regarding **participant selection**, participants were selected through convenience sampling based on the authors' academic and professional networks. The participants consisted of nine higher education members from several universities in the Computer Science department. They all had practical experience in both creating and using OER.

Their teaching experience ranged from 3 to 20 years, ensuring a diverse yet experienced team. We considered their profiles suitable for assessing the technical, economic, and social aspects of OER and for evaluating CM4OER. Despite the limitations of convenience sampling in terms of generalizability, the depth and relevance of participants' experiences enabled meaningful validation of the study's core contributions.

We developed a **semi-structured interview protocol** to guide the conversations while allowing flexibility for participants to elaborate on their experiences. We organized the protocol into three major thematic sections: (i) evaluation of the **relevance** and **clarity** of the 26 barriers identified in the literature; (ii) evaluation of the **correctness** and **completeness** of the classification of these barriers according to three aspects and the 5R's framework; and (iii) assessment of CM4OER, focusing on its **clarity**, **comprehensiveness**, and **practical usefulness**. The protocol included both closed and open-ended questions, enabling consistent data collection from participants while capturing nuanced insights. Before conducting the interviews, we conducted a pilot interview to refine the flow and clarity of the questions.

Concerning **data collection procedure**, we conducted all interviews remotely via Google Meet, each lasting approximately 50 minutes. We conducted the interviews over a 30-day period. We automatically transcribed each session using Google Meet's built-in transcription functionality. Before the interview began, participants were informed about the research goals and verbally consented to be recorded and have their conversations transcribed. They were also explicitly told that their participation was voluntary and that they could withdraw without penalty or consequence. We did not offer financial or material incentives for participation. We anonymized the transcriptions for analysis purposes. The interview protocol and transcripts of the interviews are available in the external material Available at: https://tinyurl.com/4h7j6kwa.

#### *4.1.2. Interviews Results*

The results indicate that participants agreed with the set of 26 barriers (B) and judged them to be representative, although some specific barriers required clarification.

Regarding the **understandability** of the barriers, we observed that participants (P) demonstrated comprehension of the set of barriers. P5 emphasized that "*the barriers are real problems*," and P3 noted that they "*accurately translate the challenges*" involved in adopting and sustaining OER. At the same time, some suggestions emerged during the interviews to better understand the barriers. B1 (Lack of coherent policy support across institutional and national levels undermines OER initiative) and B3 (Lack of incentives for the socialization/sharing of the OER) were perceived as ambiguous. P3 noted that it "*gives a double meaning*" and raised uncertainty about whether it refers to institutional, governmental, or international policies (e.g., UNESCO guidelines). This suggested the need to specify the level of policy influence and to clarify its effects on OER use. Regarding B3, during the interview, P1 raised questions about the type of competency this barrier refers to and who its target audience is: users, teachers, developers, or the community involved with OER. Similarly, the term *digital skills* was found to be imprecise in B2 (I.e., a Lack of digital literacy that impedes the adoption of OER). P6 proposed using *digital literacy* instead, arguing that it encompasses not only technical ability but also a critical and reflective understanding of digital practices. P1 also questioned the classification of this barrier as social, suggesting that it also aligns with the technical dimension. Both participants emphasized the importance of defining whether this competency pertains to OER authors, users, or both. Hence, we decided to maintain B2 as a barrier to the social and technical aspects. To better understand each barrier, P3 suggested that examples could be included illustrating how each barrier affects specific 5R's actions. This is an interesting suggestion, but our intention is to have a set of explicit barriers without examples.

Regarding the **completeness** of the barriers, none of the participants suggested adding or removing barriers; as P5 stated, all barriers are meaningful, though their impact varies across educational contexts.

Concerning the **correctness** of classifying barriers into aspects, participants generally agreed with the categorization into social, economic, and technical aspects but recommended specific adjustments. These refinements reflect a conceptual clarification process designed to ensure internal consistency and accuracy in representing how each barrier impacts the use of OER. In particular, B2 (Lack of digital literacy that impedes the adoption of OER) included the technical aspect. According to P1 and P7, "*its description emphasizes digital literacy rather than merely 'digital skills,' highlighting the technological proficiency required for OER adoption*". This change acknowledges that the difficulty in adopting OER is not only a social or economic issue but also intrinsically technical, as it depends on users' ability to operate digital platforms and tools effectively.

We address the issue of B3 (Lack of incentives for the socialization/sharing of OER) by considering both the technical and economic aspects, as well as the existing social dimension. P2 recognizes that "*sharing involves not only cultural and social willingness but also economic mechanisms for rewarding contribution and the technical infrastructure that enables sharing across platforms*". This reflects a broader and more accurate understanding of what sustains collaborative OER ecosystems.

B5 (Cost to OER adoption) underwent a notable reclassification. We removed the social aspects while maintaining the technical and economic dimensions. This adjustment sharpens analytical precision, isolating cost as an economic factor rather than conflating it with other difficulties. In addition, we add the technical aspect to B6 (Teachers are not interested in finding OER) while removing the economic one. P4 mentions that the *"teachers' lack of interest in finding OER is framed more as a usability and accessibility issue within digital repositories than as a problem of economic motivation*". The inclusion of the technical perspective highlights the practical challenges educators encounter when searching for and evaluating online resources.

We removed the economic aspect in B17 (Lack of knowledge about existing OER), retaining only social and technical aspects. The economic component was excluded because the challenge does not directly relate to financial constraints, as noted by P8. Finally, B18 (Lack of time to evaluate OER before using them) shifted from a social to an economic focus, while maintaining its technical nature. The time is treated as a scarce resource, reflecting opportunity cost and productivity concerns rather than interpersonal or cultural factors. This refinement aligns the classification with the implicit economic trade-offs inherent in resource evaluation and adoption.

Participants suggested other improvements in the barriers. For instance, P8 proposed grouping the barriers into thematic clusters (governance and licensing, competence and support, infrastructure and quality, and discoverability and overload) and providing short conceptual summaries for each barrier to facilitate comprehension. After analyzing this suggestion, we decided not to implement the proposed thematic clustering, as our primary objective at this stage was to validate the individual barriers independently on clarity, correctness, and relevance, rather than to reorganize them conceptually. Introducing thematic groupings could have influenced participants' interpretations, potentially biasing their judgments about specific barriers by framing them within predefined categories. Moreover, the existing classification by social, eco-nomic, and technical aspects already provides a structured analytical lens consistent with the conceptual model on barriers for OER.

In terms of **relevance** and **usefulness**, all participants acknowledged the practical value of the identified barriers. P5 observed that their applicability depends on the educational level, with some being more prominent in basic education and others in higher education. P2 emphasized that B9 (Lack of standardized metadata for OER) and B17 (Lack of knowledge about existing OER) are particularly critical barriers, as they directly hinder the discovery and reuse of OER. Collectively, these findings indicate that the barriers are relevant, actionable, and representative of real-world conditions.

The participants also evaluated **CM4OER** positively, finding it clear, coherent, and generally useful. For instance, P6 noted that the model "*fulfills its role well* " by integrating the main elements of OER. P3 stated that the diagram "*makes sense*" as it connects aspects, barriers, and mitigation actions.

Regarding **correctness** and **completeness** of CM4OER, participants proposed some improvements. P2 noted that the model does not explicitly represent how licensing influences OER use, suggesting that this relationship should be made visible in the diagram. P1 emphasized the absence of the student/teacher actor, who plays a crucial role in designing, maintaining, and enhancing OER. Including this actor would increase the model's representational accuracy. Participants also recommended explicitly linking barriers to OER actions (for example, showing how insufficient metadata affects reuse and revision, or how unclear licensing limits remix and redistribution). The barriers represent contextual constraints, while the OER actions (reuse, revise, remix, and redistribute) describe operational behaviors within the system. Connecting them directly could suggest deterministic or one-to-one causal relationships that oversimplify the multifaceted nature of challenges to using OER in practice. Moreover, the purpose of the current conceptual model is to represent the key elements and their overarching relationships rather than to model all possible interactions.

Based on these recommendations, we considered two key adjustments in our model: (i) the addition of the concept Licensing node; and (ii) inclusion of the concept *Student/Teacher* as a stakeholder connected that manages the OER.

In terms of **relevance** and **usefulness**, all participants agreed that CM4OER is a valuable tool for understanding the OER ecosystem. P6 and P7 emphasized that the model facilitates visualization of how barriers and concepts interrelate, while P4 highlighted its potential as a preventive guide for developing OER "in a more effective way." Together, these findings confirm the model's dual utility: as a theoretical synthesis and a practical framework for OER.

Therefore, the results demonstrate strong consensus among participants regarding the validity and practical significance of both the 26 barriers and the CM4OER model. The barriers were consistently described as relevant, comprehensive, and reflective of real-world challenges faced by OER educators and institutions. Importantly, no barrier was considered irrelevant or redundant. Additionally, the participants' feedback revealed opportunities for refinement, particularly in terms of clarity and the relationship between concepts.

First, some barriers, especially those related to policy and regulation, as well as literacy skills, require clarification in terms of terminology and structure. Second, the participants emphasized the need to strengthen the operational link between barriers and the 5R's framework by adding concrete examples and cause–and–effect relationships. Regarding the conceptual model, the participants' feedback indicates that CM4OER accurately captures the conceptual relationships between barriers and miti-gation actions. Overall, the findings validate both the conceptual soundness and practical relevance of the OER framework. Participants perceived it as a useful tool for identifying, categorizing, and addressing challenges for using OER in practice. The CM4OER, in particular, emerged as a practical visual and conceptual synthesis, capable of supporting both researchers and practitioners in understanding how technical, social, and economic aspects interact to affect OER.

Considering that we achieved an overall agreement among participants regarding the **understandability**, **correctness**, **completeness**, and **relevance/usefulness** of all barriers, we used them for a demonstration of how an OER can face such barriers. The final version of the barriers presented in Table 3 (Section 5) can be referred to follow the demonstration in the next section.

### *4.2. Barrier Demonstration*

As part of the evaluation, we demonstrate how a given OER faces the barriers found in this work. To do this, we selected a well-accessed OER and analyzed it. Details of this demonstration are described below.

#### *4.2.1. Selection of an OER*

Several repositories have provided OER, among which OER Commons[5] stands out as one of the most recognized and widely used repositories. Started in 2007 by the Institute for the Study of Knowledge Management in Education (ISKME)[6], OER Commons is a collaborative space for educators to share and access high-quality OER free of charge. This repository features an extensive database of OER and promotes communities of practice where educators can collaborate and innovate teaching and learning methodologies. In this repository, we selected an OER in the education field with the most views and downloads: College Success[7], with more than 67,000 views. This OER is a comprehensive guide that aims to equip college students with the necessary academic and personal skills. It covers essential topics such as time management, study techniques, communication skills, and overall well-being. It also provides strategies to overcome common challenges, such as procrastination and stress, and encourages a proactive approach to managing these issues.

#### *4.2.2. Result of analysis of Barriers Demonstration*

After thoroughly analyzing College Success, its content, format, and any associated information, we observed that it does not face 12 barriers (listed in Table 2) from a total of 26. In short, this OER is well-structured, in English, and in text/HTML format, and makes clear its licenses and usage; hence, due to its accessible and free nature, it does not face those barriers associated with OER adoption (i.e., barriers **B1**, **B2**, **B3**, **B5**, **B16**, **B24**, and **B26**), also evidenced by the significant number of accesses and downloads. Following, we discuss each barrier.

We did not find any policy regulations hindering the use of this OER (**B1**), thereby facilitating its widespread adoption and encouraging users and educators to use it. Moreover, the number of views of this OER suggests that it does not suffer from a lack of users' digital skills (**B2**). Users likely have the necessary competencies to effectively access and utilize it. We can also observe that it does not suffer from the lack of incentives for knowledge exchange (**B3**). We believe the OER's authors (from a university) were concerned about students' difficulties, which motivated them to share their knowledge. Cost-related issues are also absent in College Success (**B5**) since it is freely available, i.e., without financial restrictions that could often deter users from accessing it or educators from integrating it into the curricula. Consequently, this OER could remain an option for users and educators seeking quality materials without incurring costs. In addition, the quality of OER is critical for its use (**B16**). Analyzing College Success, we observe that it meets the quality expectations of its audience, given the number of views. Regarding the accessibility of College Success (**B24**), it is offered in an easy-access format (text/HTML) that allows users to access the content whenever needed. Regarding the lack of knowledge about OER repositories (**B26**), potential College Success users do not face this barrier once the OER is available in a well-known repository (OER Commons), as evidenced by the resource's significant number of views and downloads.

[5]https://oercommons.org/
[6]https://www.iskme.org/
[7]https://oercommons.org/courses/college-success

**Table 2.** Barriers not faced by the College Success

| ID | Barrier Description | Explanation |
|---|---|---|
| B1 | Lack of coherent policy support across institutional and national levels undermines OER initiatives | College Success is available in an open repository, indicating supportive OER policies. |
| B2 | Lack of digital literacy that impedes the adoption of OER | High number of views suggests that users possibly have the necessary digital skills. |
| B3 | Lack of incentives for the socialization/sharing of the OER | The content of College Success seems to incentive users to access it, benefiting from the knowledge that it provides. |
| B5 | Cost to OER adoption | College Success free status eliminates cost as a barrier. |
| B6 | Teachers are not interested in finding OER | High number of views could indicates that students and potentially teachers are interest in. |
| B8 | Lack of clarity about license types of OER | Successful access and usage suggest clear licensing, encouraging use. |
| B16 | Lack of quality of OER | Widespread use and popularity could imply high quality. |
| B17 | Lack of knowledge about existing OER | College Success's popularity could indicates a high awareness of this resource. |
| B19 | Lack of awareness of the benefits of OER | High access numbers reflect well-established awareness. |
| B20 | Lack of motivation to use OER | College Success suggests users are motivated to engage with it. |
| B24 | Accessibility of OER materials | The significant number of views and downloads could imply good accessibility. |
| B26 | Lack of knowledge about repositories of OER | The significant number of views and downloads could imply that many educators and students know this OER repository |

In addition, it is worth mentioning that this OER is licensed under the Creative Commons Attribution-Noncommercial-Sharealike license (**B8**). This license permits the free copying, redistribution, remixing, transformation, and building upon the material. Regarding teachers' interest in this OER (**B6**), we believe this barrier does not explicitly apply to College Success, since potential interested parties are students. At the same time, teachers could be interested in finding and recommending this OER to their students. Additionally, the awareness of College Success (**B17**), the awareness of its potential (**B19**), and the motivation to use it (**B20**) do not appear to be issues for College Success. Again, we believe the high number of views could indicate that users and educators are aware not only of College Success but also motivated to use it.

We believe the six other barriers (**B7**, **B9**, **B10**, **B22**, **B23**, and **B25**) could negatively impact the use of College Success. The existence of several OER on the same topic could also hamper the use of a given OER (**B7**), as users or educators may feel overwhelmed when choosing the most relevant one. Indeed, there are many other OER similar to College Success, leading to indecision, selection of inferior OER compared to College Success, or even non-use of OER. A lack of standardized metadata for College Success could make it difficult to find it (**B9**), hindering its visibility. Technological and infrastructural limitations could be another barrier (**B10**), as not all students may have reliable internet access or adequate devices; therefore, College Success may be of limited use to those students. The time and cost of adapting courses to include OER can be an additional barrier (**B25**). If educators feel it will take too much time or resources to adapt College Success to their curriculum, they may choose not to use it. The resistance of users and teachers to using OER developed at other institutions could be another barrier (**B22**). While College Success is valuable, it was developed by a given institution, which could lead to its non-implementation by other institutions. Similarly, the difficulty of recognizing OER from other institutions could result in OER not receiving due credit in the academic environment (**B23**). This can lead to the perception that College Success is less valid or relevant.

The remaining eight barriers (**B4**, **B11**, **B12**, **B13**, **B14**, **B15**, **B18**, and **B21**) could not able to be analyzed considering the information available for College Success. Regarding barriers that were not analyzed, we can mention social distance (**B4**), as the interaction between students and educators is crucial to learning. In this case, the OER was apparently prepared for individual students who will probably not engage with teachers about its content, so this barrier does not apply. Another example is that the pedagogical approaches adopted in OER should match the users' expectations (**B11**). It is unclear which approaches were adopted in the College Success case; therefore, this barrier was not analyzed. We also mention that the socio-cultural characteristics of OER may also hinder its adoption (**B12**). College Success was developed for students in general without explicitly addressing social or cultural issues. As the users of this OER are quite diverse, it is hard to affirm that this barrier could directly impact the use of College Success. In addition, we can highlight that the necessary digital literacy could be a critical barrier (**B13**). In the case of College Success, it would be premature to assume that all users possess the digital literacy necessary to understand the content provided by this OER; hence, we did not analyze this barrier.

The context in which a given OER is positioned can make or break it (**B14**). In this case, this barrier cannot be analyzed because College Success addresses universal competencies widely applicable in many academic and cultural contexts. We should emphasize that the barrier related to the availability of supportive tools and information is essential for downloading and using a given OER (**B15**). We did not analyze this barrier because no information was provided about what is required to use this OER. Despite this, we infer that only a browser is needed because this OER is available via HTML, and as most users have a browser, College Success would not face this barrier. Furthermore, no information was provided about previous evaluations of College Success, which hampers users and educators from recognizing the value of this OER (**B18**).

Finally, the need for training in using OER is an additional barrier that could hinder adoption (**B21**). We believe this is not the case with College Success, as it may have been prepared for individual students who required prior training. Therefore, the use of College Success could be related to its ease of access, quality, and licensing, as evidenced by the volume of accesses and downloads. Moreover, while some barriers could not be analyzed, others could hamper its long-term viability.

It is worth noting that the analysis of this OER was subjective and conducted by the authors of this paper, based on the available information. Finally, our intention with this demonstration was to illustrate how diverse aspects (i.e., barriers) of different natures can impact the use of a given OER. Therefore, this demonstration demonstrates the viability of considering barriers when analyzing a given OER.

## 5. Results

This section presents the revised classification of the identified barriers and the adjusted version of our conceptual model.

### *5.1. Barriers for OER*

Table 3 lists the 26 barriers that can hamper OER from being used and effectively reused over the years. This table lists the barrier IDs (*Bn*), the barriers themselves, and the studies that originated them. For instance, the barrier B2 "*Lack of digital literacy that impedes the adoption of OER*" came from S2, S4, and S14. The fourth, fifth, and sixth columns in Table 3 show the aspects (social, technical, and economic) of each barrier. For instance, B2 is associated with the *social* aspect because it can create disparities in access to education, leading to digital exclusion. Most barriers are associated with the technical aspect, followed by social and economic ones.

It is worth highlighting that in our analysis, we classified those barriers associated with individual aspects, such as B2 *"Lack of digital literacy that impedes the adoption of OER"* and B4 *"Perceived social distance that hampers the interaction between colleagues and learners in distributed teamwork"*, as having a social aspect. This is because social and individual aspects are often combined. Another observation is that no study identified barriers related to the environmental aspect, possibly because OER is still a relatively new research topic with opportunities for future investigation. In short, the three aspects in the context of OER can be understood as:

- **Social aspect**: It is related to the digital literacy for dealing with OER and the incentives to exchange knowledge through OER. It also refers to barriers associated with teachers' overall lack of interest in using, evaluating, and finding high-quality OER. In short, these barriers are related to the lack of digital skills in dealing with OER, as well as a lack of knowledge about existing OER for specific content in computing, such as software engineering, programming, and others;
- **Economic aspect**: It is related to the effort (time and monetary) spent to develop, select, evaluate, or adopt OER. The economic aspect puts together barriers associated with the effort of teachers searching for OER in various repositories and the financial and time costs to develop, adapt, or use OER, as well as a limited infrastructure to meet the needs; and
- **Technical aspect:** It is related to technical issues to develop, use, find, and make OER available in repositories. The technical aspect concerns the lack of knowledge about policies and technologies for implementing OER in the classroom, and understanding existing licenses and choosing OER are the two main issues. Other problems related to OER include the need to consider existing standards, such as accessibility requirements. The barriers in this aspect are related to the lack of regulation, difficulties in selecting licenses, and choosing the adequate OER for the given content. These barriers are also associated with the need for more standardization of metadata and support tools for developing and using OER.

The last five columns of Table 3 show the 5R's activities associated with each barrier. For instance, B2 *"Lack of digital literacy that impedes the adoption of OER"* can affect the OER *reuse*, while B15 *"Supporting tools and extra information required to download and use OER"* can difficult the *retain* and *remix* OER. In general, we can observe that most barriers (21 out of 26) can affect the *reuse* of OER, followed by *retain* (with 14 out of 26), while *revise*, *remix*, and *redistribute* are less recurrent with five, four, and five barriers, respectively. From the 5R's perspective analysis, other barriers could exist, for instance, for those activities that are less addressed.

We observed that barriers can affect different OER stakeholders, specifically teachers and students, as illustrated in Figure 2. In summary, we observed that most barriers reported in the literature affect teachers and relate to social and technical aspects. At the same time, barriers affecting students are the least numerous, while social barriers are more prevalent than economic and technical ones for this stakeholder. We can also say that these later aspects (economic and technical) and associated barriers from the students' perspective have drawn less attention. From this holistic perspective, readers can understand where and what the primary difficulties are in using and effectively reusing OER over the years, as well as how such barriers affect these resources and impact stakeholders. Hence, strategies and initiatives could be defined aiming to minimize those barriers. It is worth highlighting that we did not explicitly find other stakeholders, such as education institutions, governments, or countries, whose barriers could directly impact. We believe these barriers could affect these stakeholders in a second moment, but such an investigation should also be conducted further.

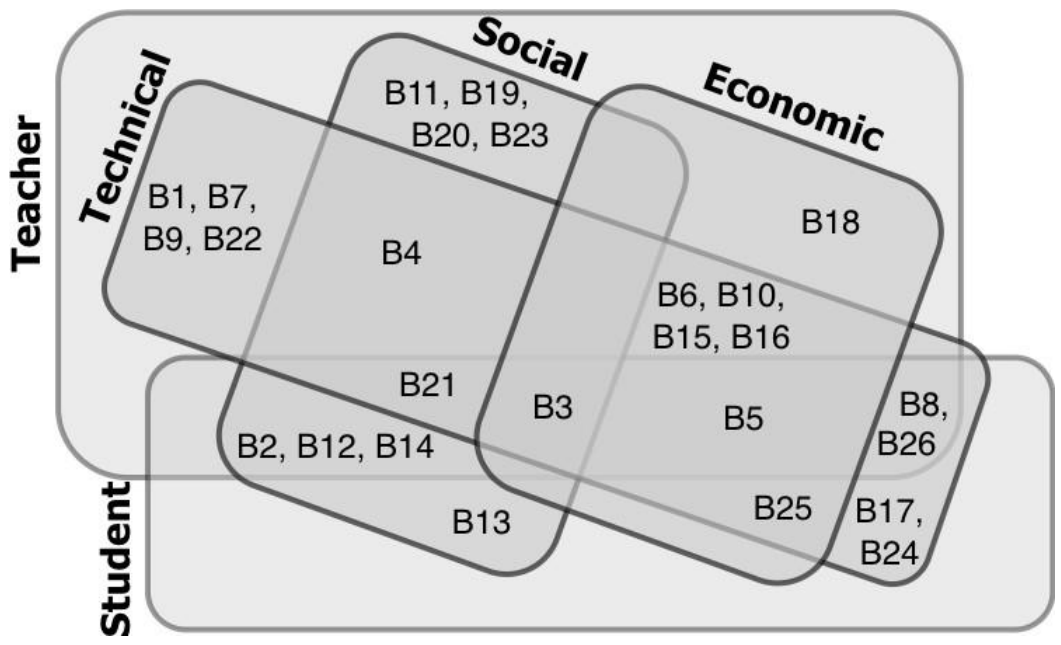

**Figure 2.** Barriers impacting the main stakeholders (teachers and students)

## *5.2. Conceptual Model and Definition of OER*

Figure 3 illustrates our conceptual model (the CM4OER), which represents the relationships between barriers and OER, highlighting the elements and factors that contribute to their long-term use and maintenance.

Within the scope of the model, several types of OER, including videos, slides, games, and books, are available in diverse formats and on multiple platforms. Each resource is associated with a specific *license*, such as Creative Commons, GPL, MIT, or Apache, that defines the permissions and restrictions on use, thereby ensuring openness and legal compliance for reuse.

**Table 3.** List of barriers to OER

| ID | Barriers | Studies | Social | Economic | Technical | Reuse | Retain | Revise | Remix | Redistribute |
|---|---|---|---|---|---|---|---|---|---|---|
| B1 | Lack of coherent policy support across institutional and national levels undermines OER initiatives | S2, S4, S15, S17, S19, S20, S21, S25, S26 | | | x | | x | | | x |
| B2 | Lack of digital literacy that impedes the adoption of OER | S2, S4, S14 | x | | | x | | | | |
| B3 | Lack of incentives for the socialization/sharing of the OER | S4, S17, S20, S25, S26 | x | x | x | x | x | | | x |
| B4 | Perceived social distance that hampers the interaction between colleagues and learners in distributed teamwork | S4 | x | | x | | x | | | x |
| B5 | Cost to OER adoption | S7, S8, S9, S15, S20 | | x | x | x | x | | | |
| B6 | Teachers are not interested in finding OER | S9 | | x | x | | x | x | | |
| B7 | Teachers face the problem of having too many choices of OER | S11 | | | x | x | | x | | |
| B8 | Lack of clarity about license types of OER | S11, S19 | | | x | x | x | | | |
| B9 | The lack of standardized metadata for OER | S11 | | | x | x | | | | |
| B10 | Technological and infrastructure limitations for developing, using, and selecting OER | S5, S14, S15, S17, S18, S22 | | x | x | x | x | x | x | |
| B11 | The pedagogical approaches adopted on OER | S5, S10, S14 | x | | | x | | | | |
| B12 | Sociocultural characteristics of the OER contents | S10, S21, S23 | x | | | x | | | | |
| B13 | The competences, abilities, and skills required of students to participate in online learning | S7, S14, S24 | x | | | x | x | | | |
| B14 | An given OER is not applicable to all contexts | S7, S14 | x | | | x | | | | |
| B15 | Supporting tools and extra information required to download and use OER | S5 | | x | x | | x | | x | |
| B16 | Lack of quality of OER | S6, S7, S12, S18, S19, S20, S21, S22, S23, S24, S25, S26 | | x | x | x | | | | |
| B17 | Lack of knowledge about existing OER | S7, S18, S19 | | | x | x | x | | | |
| B18 | Lack of time to evaluate OER before using them | S6, S18, S19, S20, S25, S26 | | x | | x | | x | | |
| B19 | Lack of awareness of the benefits of OER | S7, S16, S13, S17, S18, S19 | x | | | x | x | | | |
| B20 | Lack of motivation to use OER | S16 | x | | | x | x | | | |
| B21 | Lack of training in using OER | S16, S18, S19, S23, S25, S26 | x | | x | x | | | | |
| B22 | Resistance of teachers to using OER developed in other institutions | S13 | | | x | x | x | | | |
| B23 | Difficulty in recognizing that other institutions can develop quality OER | S13 | x | | | x | | x | x | x |
| B24 | Accesibility of OER material | S13, S21, S25, S26 | | | x | x | | | | |
| B25 | Lack of budget and effort taken for adapting courses to adopt OER | S1, S22 | | x | x | | | | x | |
| B26 | Lack of knowledge about repositories of OER | S3, S18, S19 | | | x | x | x | | | |

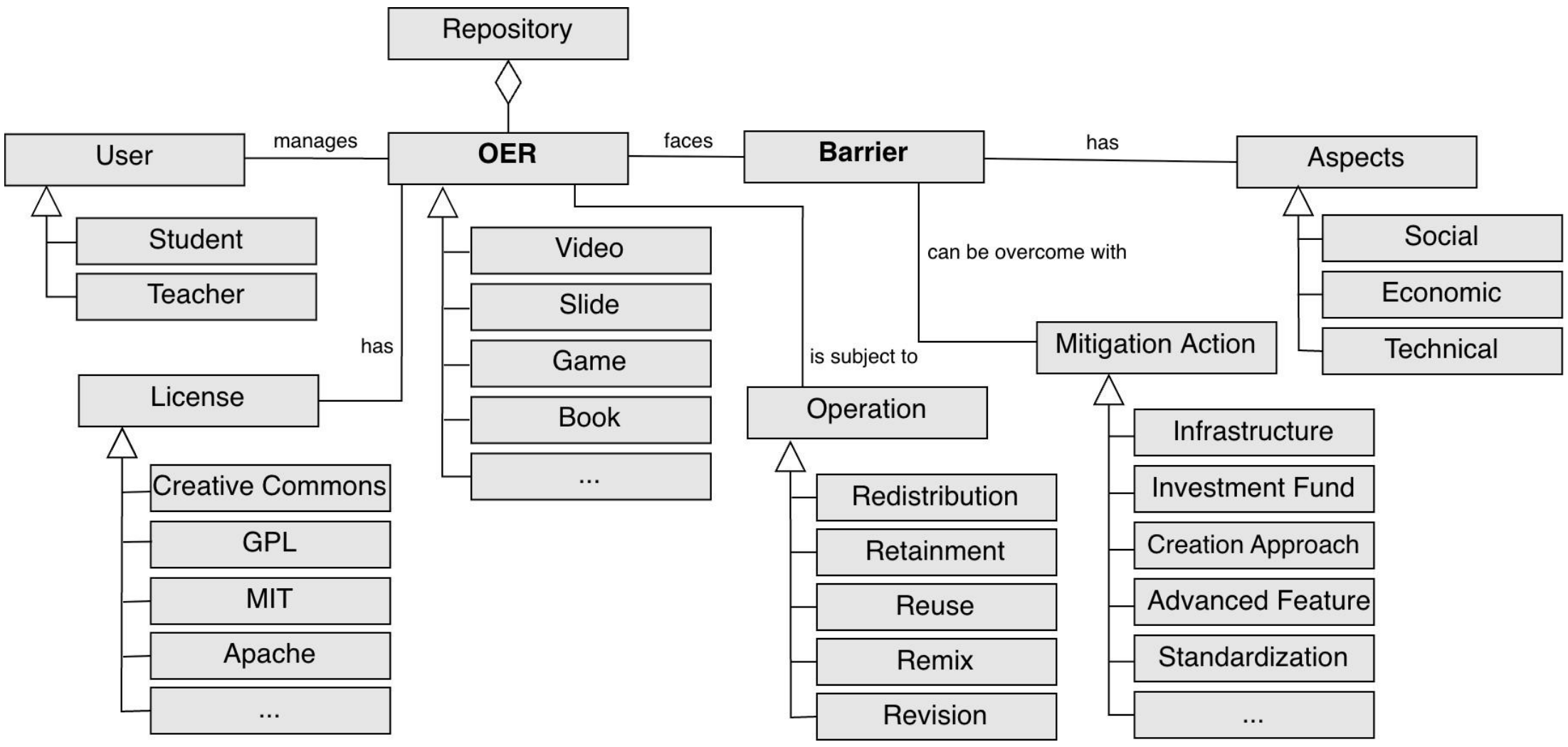

**Figure 3.** CM4OER (Conceptual Model for OER)

*Repositories* serve as the primary environment for storing and disseminating OER and are accessed by various *user* profiles, primarily students and teachers. These users serve as key agents for the use/reuse of OER, as their active participation in creating, adapting, and sharing resources extends the life cycle and leverages the educational impact of OER. However, the underuse and reuse of OER in practice are often challenged by *barriers* associated with social, economic, and technical aspects of OER. Such barriers limit the adoption, maintenance, and evolution of OER, thus requiring specific *mitigation actions*. To address these barriers, a set of mitigation actions can be adopted, including securing investment funds, enhancing technological infrastructure, developing new approaches to creating OER, implementing advanced features, and standardizing development and distribution processes. Moreover, the model considers the *operations* involved in the OER life cycle (redistribution, retainment, reuse, remix, and revision) as essential practices to promote openness and ensure the use of such resources of the educational ecosystem.

In summary, CM4OER offers an integrated view on the elements that impact the use and effective reuse of OER, highlighting key aspects, barriers, and mitigation strategies. It reinforces the importance of multidimensional and collaborative approaches to ensure the continuity, accessibility, and evolution of OER over time.

## 6. Implications for the Education Community

The findings of this work can have implications for the education community, particularly for educators, instructional designers, academic institutions, and policymakers interested in the development, adoption, and long-term stewardship of OER. By identifying and systematizing the barriers that have hindered OER, this work provides a comprehensive foundation for both immediate action and future strategic planning. Among them, we highlight:

- **Awareness and understanding of OER challenges:** This work contributes to raising awareness of the current state of OER practice by presenting one of the first consolidated and empirically grounded classifications of barriers affect-ing OER creation, use, reuse, and maintenance. By making explicit the social, organizational, technical, financial, and pedagogical constraints faced in real educational contexts, the results help the education community better understand how these barriers impact teaching practices, student learning, and institutional initiatives. This shared understanding supports more realistic expectations and informed evaluations of OER-related efforts.

- **Strategic guidance for mitigation and institutional action:** Beyond identifying challenges, classifying barriers provides concrete guidance for designing effective mitigation strategies. The findings indicate which types of interventions, such as professional development, governance structures, incentive mechanisms, community engagement, or technical infrastructure, are more likely to address specific barriers. This enables educators and institutions to adopt a more strategic, evidence-based approach to implementing OER initiatives, reducing ad hoc decisions and increasing the likelihood of successful adoption and reuse.
- **Support for long-term stewardship of OER:** Understanding the nature of OER barriers highlights the importance of adopting practices that promote the longevity of educational resources. This includes not only technical aspects (e.g., open formats, documentation, versioning, and maintenance processes) but also social and organizational mechanisms that encourage continuous improvement and shared ownership. Addressing these barriers can reduce development effort and long-term costs, improve the quality of resources, and foster more resilient OER ecosystems.
- **Cultural change and alignment with broader educational agendas:** The results emphasize the need for a cultural and institutional shift toward collaborative, open, and participatory approaches to educational resource development. Encouraging educators and institutions to move beyond a consumption-oriented mindset supports improved knowledge sharing, stronger communities of practice, and sustained engagement with OER. Furthermore, the identified barriers are closely connected to broader educational priorities, such as equity of access, digital transformation, open science, and inclusive education. Addressing them can therefore generate impacts that extend beyond OER itself, contributing to more open, equitable, and sustainable educational systems.

## 7. Discussion

While several OER have been developed to meet teachers' and students' urgent and specific needs, their widespread adoption and ease of maintenance have not been adequately addressed. In our investigation on the main reasons that prevent OER from being (re)used for an extended period, we identified 26 problems (which we denominate barriers in this work) associated with three aspects (social, economic, and technical). Analyzing the barriers enabled us to present a holistic view and better characterize the overarching problems. In addition, understanding these problems and their classification can help identify current challenges to achieving effective OER, allowing one to examine the aspects and propose ways to mitigate them.

### *7.1. Future Actions*

After conducting a thorough investigation to improve understanding of how to better address OER, we compiled potential research opportunities to advance the state of the art in OER and outlined the most urgent future actions. We organize future actions around the three aspects below.

**Social Aspect**:

- **Community Participation and Engagement:** One of the key challenges in achieving effective OER is to promote active participation and engagement of the education community. It involves encouraging teachers, students, and others to contribute to the development, improvement, and sharing of OER. Examples of solutions to these challenges include creating platforms or online communities where teachers can collaborate, exchange resources, and provide feedback on OER. Another approach is the creation of OER ambassador networks, where champions from educational institutions act as hubs to share insights among them and then spread knowledge inside their respective institutions.
- **Equity to Access:** Ensuring equitable access to OER is crucial for promoting more inclusive education. It involves addressing barriers related to geographical location, socioeconomic status, and specific learning needs. For instance, prioritizing and encouraging resources with offline access options will benefit students in areas with limited internet connectivity; also, developing OER accessible to students with disabilities can help bridge the digital divide and promote equal learning opportunities.

**Economic Aspect**:

- **Sustainable Funding Models:** Identifying viable funding models can help obtain effective OER. It can involve exploring various approaches, such as government funding, philanthropic support, public-private sector partnerships, or cost-sharing models among educational institutions. A good example is the OpenStax initiative at Rice University, which relies on grants, philanthropic donations, and institutional affiliations to sustain the development and dissemination of its open textbooks [54].
- **Viability:** Ensuring the economic viability of OER involves finding ways to support and quality assurance of OER content creation and distribution of high-quality resources while ensuring financial sustainability. It can include exploring revenue-generating models, such as offering premium features or services, providing professional development or training for OER use, or leveraging OER to drive cost savings in educational institutions by reducing reliance on costly commercial textbooks.

**Technical Aspect**:

- **Maintenance and Update:** OER requires continuous maintenance and updates to remain relevant and accurate. This involves establishing processes and mechanisms for ongoing OER content review and quality assurance. For example, creating processes and methods to support these actions can help create current, meaningful OER while also ensuring its longevity and usefulness.
- **Interoperability:** Different educational platforms, systems, and formats should be seamlessly interoperable to achieve effective OER. For instance, developing OER in standard formats, such as HTML, PDF, or EPUB, and using protocols like Learning Tools Interoperability (LTI) can ensure compatibility across various devices, learning management systems, and technologies, while also facilitating the integration and exchange of OER.

The community needs to address several future actions to leverage OER with long-term success and impact. In short, these include fostering community participation, ensuring equitable access, establishing sustainable funding models, promoting business viability, maintaining and updating content, and overcoming interoperability issues. Many of these actions are already known but have not been widely implemented. We believe these actions are the most urgent and still required.

### *7.2. Threats to Validity*

In conducting a tertiary study, we identified a set of threats to the validity of this research method. Following established guidelines in empirical software engineering [27, 53, 55, 56], we adopted a set of countermeasures to mitigate these threats, as described below.

- **Ambiguity and conceptual overlap of identified barriers.** A primary threat to construct validity concerns the risk of ambiguity when identifying and consolidating barriers from heterogeneous secondary studies. Different authors may use distinct terminology to describe similar phenomena or employ the same term with different meanings, resulting in overlapping or poorly defined constructs. To mitigate this threat, we adopted an iterative and concept-driven synthesis process. Barriers were compared based on their underlying concepts, contextual descriptions, and reported impacts rather than on terminological similarity alone. When necessary, barriers were refined, merged, or reworded to improve conceptual clarity. Additionally, each barrier was explicitly classified into social, economic, and technical categories, reinforcing construct consistency.
- **Researcher bias during barrier extraction and classification.** The qualitative nature of extracting and classifying barriers introduces the risk of researcher bias, as interpretation may be influenced by prior experience, expectations, or familiarity with OER and open education initiatives. To reduce this risk, three researchers were involved in the extraction and classification activ- ities. Decisions regarding the identification, refinement, and categorization of barriers were discussed collaboratively, and disagreements were resolved through consensus-based discussions supported by reanalysis of the original studies. Furthermore, the explicit documentation of the research protocol enhances transparency and repeatability.
- **Dependence on the quality and completeness of secondary studies.** As a tertiary study, the results inherently depend on the methodological rigor and coverage of the selected secondary studies. Limitations, biases, or omissions in those studies may propagate to the identified barrier set. To mitigate this threat, we included only peer-reviewed secondary studies retrieved from well-established digital libraries, applying explicit inclusion and exclusion criteria. When clarification was needed, primary studies referenced by the secondary studies were consulted to gain a deeper understanding of the original context of the reported barriers. This triangulation strategy is recommended to strengthen internal validity and reduce the risk of misinterpretation.
- **Generalizability of barriers across educational contexts.** Although the selected studies cover a range of educational settings, the identified barriers may not fully represent all domains, educational levels, or geographic regions. Contextual factors such as institutional culture, technological infrastructure, and national policies may influence the manifestation of barriers. To improve gen-eralizability, we included secondary studies conducted in diverse countries and educational environments. Moreover, barriers were formulated at an abstract and conceptual level, focusing on recurring structural challenges rather than context-specific symptoms. This abstraction facilitates transferability while allowing future studies to contextualize the barriers in specific domains.
- **Potential incompleteness of the identified barrier set.** Despite the comprehensive literature coverage, the identified set of barriers may not be exhaustive. Emerging technologies, evolving educational practices, and new policy frameworks may introduce additional barriers that are not yet reported in the literature. We defined a transparent and justified time frame for the literature search and complemented it by analyzing references within the selected studies. Additionally, we explicitly position the barrier set as an evolving body of knowledge rather than a definitive catalog.
- **Temporal limitation of the literature search.** Another potential threat arises from the temporal scope of the literature search, which includes secondary studies published up to 2023. More recent secondary studies or empirical evidence may identify additional barriers or refine existing ones. To mitigate this threat, we carefully analyzed the evolution of the OER research landscape over the last decade and observed a relative stability in the types of barriers reported in the literature. Recent studies tend to reinforce, refine, or recontextualize previously identified challenges rather than introduce fundamentally new categories of barriers. Moreover, the objective of this work is not to provide a time-sensitive catalog of barriers, but rather to synthesize recurring and structural challenges that hinder the long-term use of OER. Given this focus, we argue that extending the search window beyond 2023 would be unlikely to substantially alter the identified barrier set, although future updates may further strengthen or nuance the findings.

As we also conducted interviews, threats to the validity of this type of qualitative research method also exist, and, therefore, we adopted a set of countermeasures discussed below:

- **Selection of participants:** The small number of participants represents a limitation that may affect the results. Although the selected experts possessed relevant experience in both developing and using OER, the limited sample restricts the representativeness of the results. To mitigate this threat, participants were selected from various institutions and had diverse professional backgrounds, ensuring a heterogeneous set of viewpoints that could enrich the evaluation.
- **Instrumentation:** Potential bias could arise from the design of the interview questions or from unclear formulations. To address this, the interview protocol was carefully created and structured around the primary evaluation criteria (understandability, correctness, completeness, and relevance/usefulness). Additionally, a pilot interview was conducted to refine the wording and flow of the questions, ensuring that all items were comprehensible and transparent.
- **Data interpretation and analysis:** Qualitative data analysis involves a degree of subjectivity, as interpretations may differ among researchers. To minimize this threat, all interview transcripts were independently reviewed and coded by multiple authors, followed by joint discussions to reach consensus. However, given the study’s qualitative nature and small sample size, the results cannot be generalized to all OER contexts.

## 8. Conclusions

OER constitutes a crucial instrument for leveraging all educational stages, from early to higher education and corporate training. However, they have consumed considerable time and effort from their authors for creation and use, while long-term (re)use has not been widely practiced. In this scenario, this work revealed that OER still presents several barriers of different natures (social, economic, and technical), hampering their effective, long-term (re)use. In summary, these barriers are often caused by a lack of understanding of OER's benefits and by the absence of incentives to create and disseminate high-quality, cost-effective, and long-lasting OER. In particular, important stakeholders, including governments, educational institutions, and instructors, have not widely promoted OER, despite the existence of important initiatives from high-ranking universities worldwide.

This work aims to raise awareness about these barriers and highlight urgent actions to overcome them. We also believe that OER can only become widely used and effectively reused over time when key stakeholders shift their mindset, invest efforts in this direction, create communities of practice and collaboration networks, and foster a culture of sharing and collaboration in education, thereby reaping the benefits of OER for society.

Future actions were intentionally separated from the traditional "future work" section. While the conclusion summarizes the paper's main contributions, future work typically highlights open research opportunities. The discussion section (section 7), however, emphasizes actionable directions that can be directly derived from our findings. These actions were organized around the three core aspects considered in this work (social, technical, and economic), enabling a more structured and targeted discussion. This separation enables a clearer distinction between long-term research challenges and short- to medium-term actions that can support the practical adoption, long-term use, and evolution of OER. Finally, after overcoming such significant barriers, the community could harness the full potential of OER, ensuring accessibility, quality, and equity in education worldwide.

## Disclosure statement

No potential conflict of interest was reported by the author(s).

## Funding

The authors thank CNPq (313245/2021-5), FAPEMIG (APQ-00743-22), and FAPESP (2023/00488-5) for the financial support for this study.

## References

[1] S. Criollo-C, A. Guerrero-Arias, Á. Jaramillo-Alcázar, and S. Luján-Mora, *Mobile learning technologies for education: Benefits and pending issues*, Applied Sciences 11 (2021), pp. 1–17.

[2] M. Prensky, *Digital natives, digital immigrants*, Gifted (2005), pp. 29–31.

[3] O.B. Adedoyin and E. Soykan, *Covid-19 pandemic and online learning: the challenges and opportunities*, Interactive Learning Environments 31 (2020), pp. 863–875.

[4] W. Deus and E. Barbosa, *A systematic mapping of the classification of open educational resources for computer science education in digital sources*, IEEE Transactions on Education 65 (2021), pp. 450–460.
[5] W. Admiraal, *A typology of educators using open educational resources for teaching*, International Journal on Studies in Education (IJonSE) 4 (2022), pp. 1–23.
[6] R.K. Moloo, K.K. Khedo, and T.V. Prabhakar, *An audio mooc framework for the digital inclusion of low literate people in the distance education process*, Universal Access in the Information Society 24 (2025), pp. 313–337.
[7] R. Kimmons, *Expansive openness in teacher practice*, Teachers College Record 118 (2016), pp. 1–34.
[8] B. Chae and M. Jenkins, *A qualitative investigation of faculty open educational resource usage in the washington community and technical college system: Models for support and implementation*, Washington State Board for Community & Technical Colleges (2015), pp. 1–38.
[9] S. Park and K. McLeod, *Multimedia open educational resources in mathematics for high school students with learning disabilities*, Journal of Computers in Mathematics and Science Teaching 37 (2018), pp. 131–153.
[10] S. United Nations Educational and C. Organization, *Forum on the impact of open courseware for higher education in developing countries*, final report (2002).
[11] J. Hyl´en, *Open educational resources: Opportunities and challenges*, Proceedings of Open Education (2006), pp. 49–63.
[12] B.K. Swain and R.K. Pathak, *Benefits and challenges of using oer in higher education: a pragmatic review*, Discover Education 3 (2024), p. 81.
[13] M. King, M. Pegrum, and M. Forsey, *Moocs and oer in the global south: Problems and potential*, The International Review of Research in Open and Distributed Learning 19 (2018).
[14] D. Otto, N. Schroeder, D. Diekmann, and P. Sander, *Trends and gaps in empirical research on open educational resources (OER): A systematic mapping of the literature from 2015 to 2019*, Contemporary Educational Technology 13 (2021), p. ep325.
[15] C. Carvalho, M. Rodr´ıguez, P. Escudeiro, M. Nistal, *et al.*, *Sustainability of open educational resources: The eCity case*, in *International Symposium on Computers in Education (SIIE)*. 2016, pp. 1–6.
[16] S. Hoosen and N. Butcher, *Understanding the impact of OER: Achievements and challenges*, UNESCO Institute for Information Technologies in Education: Moscow, Russia (2019).
[17] D. Orr, J. Neumann, and J. Muuß-Merholz, *German OER practices and policy—from bottom-up to top-down initiatives*, UNESCO Institute for Information Technologies in Education, Moscow (2017).
[18] M.S. Ramirez-Montoya, *Challenges for open education with educational innovation: A systematic literature review*, Sustainability 12 (2020), p. 7053.
[19] B. Penzenstadler, V. Bauer, C. Calero, and X. Franch, *Sustainability in software engineering: A systematic literature review*, in *16th International Conference on Evaluation Assessment in Software Engineering (EASE)*. 2012, pp. 32–41.
[20] A. Królak and P. Zajac, *Analysis of the accessibility of selected massive open online courses (moocs) for users with disabilities*, Universal Access in the Information Society 23 (2024), pp. 191–202.
[21] A. Tlili, F. Nascimbeni, D. Burgos, X. Zhang, R. Huang, and T.W. Chang, *The evolution of sustainability models for open educational resources: Insights from the literature and experts*, Interactive Learning Environments (2020), pp. 1–16.
[22] A. Bozkurt, S. Koseoglu, and L. Singh, *An analysis of peer reviewed publications on openness in education in half a century: Trends and patterns in the open hemisphere*, Australasian Journal of Educational Technology 35 (2019), pp. 78–97.
[23] O. Zawacki-Richter, D. Conrad, A. Bozkurt, C.H. Aydin, S. Bedenlier, I. Jung, J. Stöter, G. Veletsianos, L.M. Blaschke, M. Bond, *et al.*, *Elements of open education: An invitation*

*to future research*, International Review of Research in Open and Distributed Learning 21 (2020), pp. 319–334.

[24] S. Johnstone, *Forum on the impact of open courseware for higher education in developing countries–final report*, Education Quarterly 3 (2005), pp. 15–18.

[25] J. Hilton, *Open educational resources and college textbook choices: A review of research on efficacy and perceptions*, Educational technology research and development 64 (2016), pp. 573–590.

[26] D. Wiley and J.L. Hilton III, *Defining OER-enabled pedagogy*, The International Review of Research in Open and Distributed Learning 19 (2018), pp. 133–147.

[27] B.A. Kitchenham, D. Budgen, and P. Brereton, *Evidence-based software engineering and systematic reviews*, Vol. 4, CRC press, Boca Raton, USA, 2015.

[28] A. Murphy, *Open educational practices in higher education: Institutional adoption and challenges*, Distance Education 34 (2013), pp. 201–217.

[29] C. Cobo, *Exploration of open educational resources in non-english speaking communities*, International Review of Research in Open and Distributed Learning 14 (2013), pp. 106–128.

[30] J.S. Mtebe and R. Raisamo, *Investigating perceived barriers to the use of open educational resources in higher education in Tanzania*, International Review of Research in Open and Distributed Learning 15 (2014), pp. 43–66.

[31] J. Stoffregen, J.M. Pawlowski, and H. Pirkkalainen, *A barrier framework for open e-learning in public administrations*, Computers in Human Behavior 51 (2015), pp. 674–684.

[32] K. Clements, J. Pawlowski, and N. Manouselis, *Open educational resources repositories literature review–towards a comprehensive quality approaches framework*, Computers in Human Behavior 51 (2015), pp. 1098–1106.

[33] O.M. Belikov and R. Bodily, *Incentives and barriers to OER adoption: A qualitative analysis of faculty perceptions*, Open Praxis 8 (2016), pp. 235–246.

[34] C. Hassall and D.I. Lewis, *Institutional and technological barriers to the use of open educational resources (OERs) in physiology and medical education*, Advances in Physiology Education 41 (2017), pp. 77–81.

[35] V. Clinton, *Cost, outcomes, use, and perceptions of open educational resources in psychology: A narrative review of the literature*, Psychology Learning & Teaching 18 (2019), pp. 4–20.

[36] M. King, *Doing MOOCs in Dili: Studying online learner behaviour in the Global South*, in *Proceedings of Work in Progress Papers of the Research, Experience and Business Tracks at EMOOCs*, Vol. 2356. 2019, pp. 54–59.

[37] T. Heck, S. Kullmann, J. Hiebl, N. Schröder, D. Otto, and P. Sander, *Designing open informational ecosystems on the concept of open educational resources*, Open Education Studies 2 (2020), pp. 252–264.

[38] T. Luo, K. Hostetler, C. Freeman, and J. Stefaniak, *The power of open: Benefits, barriers, and strategies for integration of open educational resources*, Open Learning: The Journal of Open, Distance and e-Learning 35 (2020), pp. 140–158.

[39] X. Meng, C. Cui, and X. Wang, *Looking Back Before We Move Forward: A Systematic Review of Research on Open Educational Resources*, in *9th International Conference of Educational Innovation through Technology (EITT)*. 2020, pp. 92–96.

[40] V. Truong, T. Denison, and C.M. Stracke, *Developing institutional open educational resource repositories in vietnam: Opportunities and challenges*, International Review of Research in Open and Distributed Learning 22 (2021), pp. 109–124.

[41] M. Oliveira, L. Paschoal, and E. Barbosa, *Quality models and quality attributes for open educational resources: a systematic mapping*, in *IEEE Frontiers in Education Conference (FIE)*. 2021, pp. 1–9.

[42] H.M. Adil, S. Ali, M. Sultan, M. Ashiq, and M. Rafiq, *Open education resources' benefits and challenges in the academic world: a systematic review*, Global Knowledge, Memory and Communication 73 (2022), pp. 274–291.

[43] M. Mi´cunovi´c, S. Rako, and K. Feldvari, *Open educational resources (oers) at european*

*higher education institutions in the field of library and information science during covid-19 pandemic*, Publications 11 (2023), p. 38.

[44] A.M. Mullens and B. Hoffman, *The affordability solution: A systematic review of open educational resources*, Educational Psychology Review 35 (2023), p. 72.

[45] F. Iniesto and C. Bossu, *Equity, diversity, and inclusion in open education: A systematic literature review*, Distance Education 44 (2023), pp. 694–711.

[46] L. Sousa, L. Pedro, and C. Santos, *A systematic review of systematic reviews on open educational resources: An analysis of the legal and technical openness*, International Review of Research in Open and Distributed Learning 24 (2023), pp. 18–33.

[47] F. Khalid, M. Wu, D.K. Ting, B. Thoma, M.R. Haas, M.J. Brenner, Y. Yilmaz, Y.M. Kim, and T.M. Chan, *Guidelines: the do's, don'ts and don't knows of creating open educational resources*, Perspectives on Medical Education 12 (2023), p. 25.

[48] M. dos Santos Soares, G.K. Kakinohana, M.I. Cagnin, K. dos Santos Silva, A. de Lima Fontão, A.P. Freire, and D.M.B. Paiva, *Pedagogical and accessibility guidelines for open educational resources focusing on blind students*, in *Simpósio Brasileiro sobre Fatores Humanos em Sistemas Computacionais (IHC)*. SBC, 2024, pp. 714–726.

[49] D.A. Aksoy, E. Kur¸sun, and O. Zawacki-Richter, *Factors affecting the sustainability of open educational resource initiatives in higher education: A systematic review*, Review of Education 13 (2025), p. e70029.

[50] S. Mishra, *A review of reviews on open educational resources*, Open Praxis 17 (2025), pp. 305–325.

[51] S. Kvale and S. Brinkmann, *Interviews: Learning the craft of qualitative research interviewing*, Sage, Los Angeles, USA, 2009.

[52] M.D. Myers and M. Newman, *The qualitative interview in is research: Examining the craft*, Information and organization 17 (2007), pp. 2–26.

[53] P. Runeson and M. Höst, *Guidelines for conducting and reporting case study research in software engineering*, Empirical software engineering 14 (2009), pp. 131–164.

[54] N. Finkbeiner, *Learn about openstax textbooks* (2019). Available at https://engagedscholarship.csuohio.edu/oa/openstax2019/openstax-day/1/, access in 24/12/2025.

[55] A. Ampatzoglou, S. Bibi, P. Avgeriou, M. Verbeek, and A. Chatzigeorgiou, *Identifying, categorizing and mitigating threats to validity in software engineering secondary studies*, Information and Software Technology 106 (2019), pp. 201–230.

[56] C. Wohlin, P. Runeson, M. Höst, M.C. Ohlsson, B. Regnell, and A. Wessl´en, *Experimentation in Software Engineering*, Springer, 2012.